\documentclass[cits]{PoS}
\usepackage{amsmath,amssymb,mathptmx,slashed}
\newcommand{\nnnlo}{{NNNLO}}
\newcommand{\ve}{{\epsilon}}
\newcommand{\ep}{{\varepsilon}}
\newcommand{\Lx}{\left(}
\newcommand{\Rx}{\right)}

\newcommand{\order}[1]{{\cal O}(#1)}
\newcommand{\aspi} {{\Lx\frac{\alpha_s}{\pi}\Rx}}
\newcommand{\abbrev}{\small}
\newcommand{\REDUZE}{{\abbrev REDUZE2}}
\newcommand{\QGRAF}{{\abbrev QGRAF}}

\newcommand{\FORM}{{\abbrev FORM}}

\newcommand{\hypgeo}[4]{{\vphantom{F}}_2F_1\Lx{#1},{\,#2};{\,#3};{\,#4}\Rx}

\newcommand{\cg}{{c_\Gamma}}
\newcommand{\sos}[2]{\frac{s_{#1}}{s_{#2}}}
\newcommand{\xb}{\bar{x}\,}
\newcommand{\yb}{\bar{y}\,}

\title{One-loop Single Real Emission Contributions to Inclusive Higgs Production at NNNLO}

\ShortTitle{One-loop Single Real Emission Contributions to Inclusive Higgs Production at NNNLO}

\author{\speaker{W. Kilgore}%
 \\
        Physics Department, Brookhaven National Laboratory Upton, New York 11973, U.S.A.\\
        E-mail: \email{kilgore@bnl.gov}}

\abstract{I discuss the contributions of the one-loop
  single-real-emission amplitudes, $gg\to H g$, $qg\to H q$, etc. to
  inclusive Higgs boson production through
  next-t0-next-to-next-to-leading order in the strong coupling.}

\FullConference{Loops and Legs in Quantum Field Theory\\
                 27 April 2014 - 02 May 2014\\
                 Weimar, Germany}

\begin{document}

\section{The Discovery of the Higgs Boson}
The most important result from the early runs at the LHC has been the
discovery of a $126$ GeV scalar boson that my be the long-sought Higgs
boson of the Standard Model.  While the measurements of its couplings
still have large uncertainties, the gross features look very much like
what is expected for the Higgs.  The identification of this particle,
the determination of whether it is the SM Higgs boson, a component of
an extended symmetry breaking sector or even an impostor, is the most
important task in our field today.

In order to make this identification, we need to measure the ``Higgs''
as thoroughly as possible.  We need to measure the mass, the width, the
cross section, and the couplings.  We need to look for more ``Higgs''
bosons and for other new particles that might be connected to the
``Higgs''.  These are difficult tasks and will require a great deal of
data at the full energy of the LHC.

One of the simplest observables associated with the Higgs, the cross
section, is actually difficult to take advantage of.  The reason for
this is that the theoretical uncertainty in the cross section is very
large.  This is the case even though the cross section has been
computed at next-to-next-to-leading
order~\cite{Harlander:2002wh,Anastasiou:2002yz,Ravindran:2003um} and
resummed to next-to-next-to-leading log
accuracy~\cite{Catani:2003zt,Moch:2005ky}.

One of the main sources of uncertainty in the gluon fusion cross
section comes from the scale dependence of the partonic cross
section.  This uncertainty can be addressed by computing the cross
section at higher order in $\alpha_s$.  This means performing the
calculation at next-to-next-to-next-to-leading order (\nnnlo).

\section{Inclusive Higgs Production at \nnnlo}
There are many contributions to inclusive Higgs production at \nnnlo,
and all have been computed in the threshold approximation: Virtual
contributions through three
loops~\cite{Lee:2010cga,Baikov:2009bg,Gehrmann:2010ue,Gehrmann:2010tu},
one-loop single real emission
squared~\cite{Anastasiou:2013mca,Kilgore:2013gba}, two-loop single
real emission~\cite{Anastasiou:2014vaa}, one-loop double real
emission~\cite{Anastasiou:2014vaa,Li:2014bfa} and triple real emission
at tree level~\cite{Anastasiou:2013srw}.  Virtual corrections only
contribute at threshold, so that term is known completely, as an
expansion in the dimensional parameter $\epsilon$.  The full kinematic
dependence of the squared one-loop single real emission contrbution,
the subject of this talk, has also been computed as an expansion in
$\epsilon$~\cite{Anastasiou:2013mca,Kilgore:2013gba}. Among the
reasons that full kinematic dependence of this term can be computed
are that the one-loop amplitudes are known in closed analytic form and
because the phase space element for single-real emission is
particularly simple.

\subsection{The Heavy Top Effective Theory}
The Higgs boson couples to mass, therefore it does not couple directly
to massage gauge bosons like gluons and photons.  Instead, such
particles have indirect couplings to the Higgs through heavy particle
loops.  The interaction between gluons and the Higgs is dominated by
the top quark, while photons couple through both top and $W$ boson
loops.  Because the top pair production threshold is much heavier than
the Higgs, one can form an effective Lagrangian for Higgs -- gluon
interactions by integrating out the top
quark~\cite{Shifman:1979eb,Voloshin:1986tc,Vainshtein:1980ea} for the
Higgs-gluon interaction:
\begin{equation}
{\cal L}_{\rm eff} 
= -\frac{H}{4v} C_1(\alpha_s)\,G^a_{\mu\nu} G^{a\,\mu\nu}\,,
\label{eqn::leff}
\end{equation}
where $C_1$ is the Wilson coefficient, known to $\order{\alpha_s^4}$
\cite{Chetyrkin:1998un,Chetyrkin:1997iv,Schroder:2005hy,Chetyrkin:2005ia},
and $G^a_{\mu\nu}$ is the gluon field strength tensor.  Using the
effective Lagrangian greatly simplifies calculations as it transforms
massive top quark loops in to point-like vertices.

\subsection{One-loop Single Real Emission}
The amplitude for single-real emission, computed at any order, can be
written in terms of a small number of gauge invariant tensors.  There
are four gauge invariant tensor structures for $Hggg$
amplitudes~\cite{Ellis:1988xu,Gehrmann:2011aa} and only two
structures~\cite{Gehrmann:2011aa} for $Hq\overline{q}g$ amplitudes,
\begin{equation*}
\begin{split}
   {\cal M}(H;g_1,g_2,g_3) &= \frac{g}{v}\,C_1(\alpha_s)\,f^{ijk}
   \ve^i_{1\mu}\ve^j_{2\nu}\ve^k_{3\rho}\sum_{n=0}^3A_n\,
  {\cal Y}_n^{\mu\nu\rho}\,,\\
   {\cal M}(H;g,q, \overline{q}) =\ &
      i\frac{g}{v}\,C_1(\alpha_s)\Lx T^g\Rx^{\bar\imath}_j\ve_\mu(p_g)
      \Lx B_1\,{\cal X}^\mu_1 + B_2\,{\cal X}^\mu_2\Rx\,.
\end{split}
\end{equation*}
The coefficient of each tensor has an expansion in $\alpha_s$ of the
form
\begin{equation}
  A_i = A_i^{(0)} + \aspi\,A_i^{(1)} + \aspi^2\,A_i^{(2)} + \ldots\,,
\end{equation}
and the same for the $B_i$.  I have computed the amplitudes in the
following manner: the Feynman diagrams were generated using
\QGRAF~\cite{Nogueira:1993ex}; they were contracted with the
projectors onto the gauge-invariant tensors and the Feynman rules were
implemented using a \FORM~\cite{Vermaseren:2000nd} program.  For the
one-loop amplitudes, the resulting expressions were reduced to loop
master integrals with the program \REDUZE~\cite{vonManteuffel:2012np}.
The reduced expressions were put back into the FORM\ program and the
master integrals were evaluated to produce the final expressions.

There are two loop master integrals that appear in these amplitudes,
the bubble and the single-mass box.  Both are known in closed analytic
form for arbitrary kinematics.  In a frame where $s_{12} > 0$,
$s_{23}, s_{31} < 0$
\begin{equation}
\begin{split}
\label{eqn:oneloopmasters}
{\cal I}^{(1)}_2(Q^2) &= 
    \frac{i\,\cg}{\ep\,(1-2\,\ep)}\left(\frac{\mu^2}{-Q^2}\right)^\ep\\
{\cal I}^{(1)}_4(s_{12},s_{23};M_H^2) &=
    \frac{2\,i\cg}{s_{12}\,s_{23}}\frac{1}{\ep^2}
    \left[\left(\frac{\mu^2}{-s_{12}}\right)^\ep
          \hypgeo{1}{-\ep}{1-\ep}{-\sos{31}{23}}\right.\\
   &
        + \left(\frac{\mu^2}{-s_{23}}\right)^\ep
          \hypgeo{1}{-\ep}{1-\ep}{-\sos{31}{12}}\\
   &\left.
        - \left(\frac{\mu^2}{-M_H^2}\right)^\ep
          \hypgeo{1}{-\ep}{1-\ep}{-\frac{M_H^2\,s_{31}}{s_{12}\,s_{23}}}
    \right]\\
{\cal I}^{(1)}_4(s_{23},s_{31};M_H^2) &=
    \frac{2\,i\,\cg}{s_{23}\,s_{31}}\frac{1}{\ep^2}
    \left[\left(\frac{\mu^2}{-s_{12}}\right)^{-\ep}
      \left(\frac{\mu^2}{-s_{23}}\right)^\ep\left(\frac{\mu^2}{-s_{31}}\right)^\ep
      \Gamma(1-\ep)\Gamma(1+\ep)\right.\\
   &+ \left(\frac{\mu^2}{-s_{23}}\right)^\ep\Lx1-
          \hypgeo{1}{\ep}{1+\ep}{-\sos{31}{12}}\Rx\\
   &
        + \left(\frac{\mu^2}{-s_{31}}\right)^\ep\Lx1-
          \hypgeo{1}{\ep}{1+\ep}{-\sos{23}{12}}\Rx\\
   &\left.
        - \left(\frac{\mu^2}{-M_H^2}\right)^\ep\Lx1-
          \hypgeo{1}{\ep}{1+\ep}{-\frac{s_{23}\,s_{31}}{s_{12}\,M_H^2}}\Rx
    \right]\,,\\
\end{split}
\end{equation}
where
\begin{equation}
\begin{split}
 \label{eq::cgamma}\cg = \frac{\Gamma(1+\ep)\Gamma^2(1-\ep)}
    {\Lx4\pi\Rx^{2-\ep}\Gamma(1-2\ep)}.
\end{split}
\end{equation}

\subsection{Squared amplitudes and Phase Space Integration}
\label{sec:SRE}
The partonic cross section is computed by squaring the
amplitudes and integrating over phase space
\begin{equation}
\sigma = \frac{1}{2\,s_{12}}\,d(LIPS)\,\frac{1}{\mathbb{S}}
  \sum_{\rm spin/color} \,\left|{\cal M}\right|^2\,,
\end{equation}
where the factor of $1/(2\,s_{12})$ is the flux factor, $d(LIPS)$
represents Lorentz invariant phase space and the factor $1/\mathbb{S}$
represents the averaging over initial state spins and colors.  The
element of Lorentz invariant phase space is
\begin{equation}
d(LIPS) = \frac{1}{8\,\pi}\Lx\frac{4\,\pi\,\mu^2}{s_{12}}\Rx^\ep\,\
   \frac{\Lx s_{23}\,s_{31}\Rx^\ep}{\Gamma(1-\ep)}\,ds_{23}\,.
\end{equation}
Defining $s_{12} = \hat{s}$ to be the parton CM energy squared, I
introduce the dimensionless parameters $x = M_H^2/\hat{s}$,
$\xb = 1-x$, and $y = \frac{1}{2}\Lx1-\cos\,\theta^*\Rx$, $\yb = 1-y$, where
$\theta^*$ is the scattering angle in the CM frame,
\begin{equation}
\begin{split}
  s_{12} &= \hat{s}\,,\hskip 50pt  M_H^2 = x\,\hat{s}\,,\\
  s_{23} &= \xb\,y\,\hat{s}\,,\hskip 37pt  s_{31} = \xb\,\yb\,\hat{s}\,.
\end{split}
\end{equation}
In terms of these variables, the element of phase space is
\begin{equation}
  d(LIPS) = \frac{1}{8\,\pi}\Lx\frac{4\,\pi^2\,\mu^2}{\hat{s}}\Rx^\ep
    \frac{1}{\Gamma(1-\ep)}\,\xb^{1-2\ep}\,y^{-\ep}\,\yb^{-\ep}\,dy\,.
\end{equation}
$\xb$ is the threshold parameter, and is a measure of excess or
kinetic energy in the scattering process, beyond that which is needed to
produce a Higgs boson at rest.  The kinematically available region in
$x$ and $y$ space is $M_H^2/s < x < 1$ and $0 < y < 1$, where
$s$ is the hadronic (not partonic) CM energy.  Clearly, $0 < \xb\ <
1-M_H^2/s$ and $0<\yb\ <1$.

\subsection{Phase Space Master Integrals}
The resulting expression for the partonic cross section consists of a
large number of phase space integrals, often with complicated
integrands involving the products of two hypergoemtric functions.  To
simplify the integration, I employ an integration-by-parts style
reduction on the phase space integrals.  Under the assumption that any
term in the integrand can be expressed in the form
$f(\xb)\,y^\alpha\,\yb^\beta$, where $\alpha$ and $\beta$ are not
negative integers (which is ensured by dimensional regularization), I
can use the fact that
\begin{equation}
  \int_0^1\ dy\ \frac{d}{dy} f(\xb)\,y^\alpha\,\yb^\beta = 0
\end{equation}
to derive relations among various phase space integrals.  Eventually,
I am able to reduce the problem to that of solving for 24 phase space
master integrals.

I compute the phase space master integrals by performing an extended
threshold expansion of some 120 terms for each master integral.  I
then map the expansions onto a set of harmonic polylogarithms and
thereby obtain the results for the master integrals in closed analytic
form.  

Finally, I substitute the values of the master integrals into the full
expression to obtain the squared one-loop single real emission
contribution to inclusive Higgs production cross section at \nnnlo.  I
find complete analytic agreement with Ref.~\cite{Anastasiou:2013mca}.
The full result is too lengthy to report here, but can be obtained
from the supplementary material attached to the journal article at
http://link.aps.org/supplemental/10.1103/PhysRevD.89.073008.

\section{Conclusions}
I have computed the contributions of one-loop single-real-emission
amplitudes to inclusive Higgs boson production at \nnnlo.  Though a
complicated calculation, this is but a portion of the full
\nnnlo\ result.  I have computed this contribution as an extended
threshold expansion, obtaining enough terms to invert the series and
determine the closed functional form through order $\ep^1$.

\paragraph*{Acknowledgments:}
This research was supported by the U.S.~Department of Energy under
Contract No.~DE-AC02-98CH10886.

\providecommand{\href}[2]{#2}\begingroup\raggedright\endgroup


\begin{thebibliography}{10}

\bibitem{Harlander:2002wh}
R.~V. Harlander and W.~B. Kilgore,  {\em Next-to-next-to-leading order {Higgs}
  production at hadron colliders}, {\em Phys. Rev. Lett.} {\bf 88} (2002)
  201801,
\href{http://arXiv.org/abs/hep-ph/0201206}{{\tt[hep-ph/0201206]}}.

\bibitem{Anastasiou:2002yz}
C.~Anastasiou and K.~Melnikov,  {\em {Higgs} boson production at hadron
  colliders in {NNLO} {QCD}}, {\em Nucl. Phys.} {\bf B646} (2002) 220--256,
\href{http://arXiv.org/abs/hep-ph/0207004}{{\tt[hep-ph/0207004]}}.

\bibitem{Ravindran:2003um}
V.~Ravindran, J.~Smith, and W.~L. Van~Neerven,  {\em {NNLO} corrections to
  the total cross section for {Higgs} boson production  in hadron hadron
  collisions}, {\em Nucl. Phys.} {\bf B665} (2003) 325--366,
\href{http://www.arXiv.org/abs/hep-ph/0302135}{{\tt[hep-ph/0302135]}}.

\bibitem{Catani:2003zt}
S.~Catani, D.~de~Florian, M.~Grazzini, and P.~Nason,  {\em Soft-gluon
  resummation for Higgs boson production at hadron colliders}, {\em JHEP} {\bf
  07} (2003) 028,
\href{http://www.arXiv.org/abs/hep-ph/0306211}{{\tt[hep-ph/0306211]}}.

\bibitem{Moch:2005ky}
S.~Moch and A.~Vogt,  {\em {Higher-order soft corrections to lepton pair and
  Higgs boson production}}, {\em Phys.Lett.} {\bf B631} (2005) 48--57,
\href{http://www.arXiv.org/abs/hep-ph/0508265}{{\tt[hep-ph/0508265]}}.

\bibitem{Lee:2010cga}
R.~Lee, A.~Smirnov, and V.~Smirnov,  {\em {Analytic Results for Massless
  Three-Loop Form Factors}}, {\em JHEP} {\bf 1004} (2010) 020,
\href{http://www.arXiv.org/abs/1001.2887}{{\tt[1001.2887]}}.

\bibitem{Baikov:2009bg}
P.~Baikov, K.~Chetyrkin, A.~Smirnov, V.~Smirnov, and M.~Steinhauser,  {\em
  {Quark and gluon form factors to three loops}}, {\em Phys.Rev.Lett.} {\bf
  102} (2009) 212002,
\href{http://www.arXiv.org/abs/0902.3519}{{\tt[0902.3519]}}.

\bibitem{Gehrmann:2010ue}
T.~Gehrmann, E.~Glover, T.~Huber, N.~Ikizlerli, and C.~Studerus,  {\em
  {Calculation of the quark and gluon form factors to three loops in QCD}},
  {\em JHEP} {\bf 1006} (2010) 094,
\href{http://www.arXiv.org/abs/1004.3653}{{\tt[1004.3653]}}.

\bibitem{Gehrmann:2010tu}
T.~Gehrmann, E.~Glover, T.~Huber, N.~Ikizlerli, and C.~Studerus,  {\em {The
  quark and gluon form factors to three loops in QCD through to
  O($\epsilon^2$)}}, {\em JHEP} {\bf 1011} (2010) 102,
\href{http://www.arXiv.org/abs/1010.4478}{{\tt[1010.4478]}}.

\bibitem{Anastasiou:2013mca}
C.~Anastasiou, C.~Duhr, F.~Dulat, F.~Herzog, and B.~Mistlberger,  {\em
  {Real-virtual contributions to the inclusive Higgs cross-section at
  $N^3LO$}}, {\em JHEP} {\bf 1312} (2013) 088,
\href{http://www.arXiv.org/abs/1311.1425}{{\tt[1311.1425]}}.

\bibitem{Kilgore:2013gba}
W.~B. Kilgore,  {\em {One-Loop Single-Real-Emission Contributions to $pp\to H +
  X$ at Next-to-Next-to-Next-to-Leading Order}}, {\em Phys.Rev.} {\bf D89}
  (2014) 073008,
\href{http://www.arXiv.org/abs/1312.1296}{{\tt[1312.1296]}}.

\bibitem{Anastasiou:2014vaa}
C.~Anastasiou, C.~Duhr, F.~Dulat, E.~Furlan, T.~Gehrmann, {\em et al.},  {\em
  {Higgs boson gluon-fusion production at threshold in N3LO QCD}},
\href{http://www.arXiv.org/abs/1403.4616}{{\tt[1403.4616]}}.

\bibitem{Li:2014bfa}
Y.~Li, A.~von Manteuffel, R.~M. Schabinger, and H.~X. Zhu,  {\em {N$^3$LO Higgs
  and Drell-Yan production at threshold: the one-loop two-emission
  contribution}},
\href{http://www.arXiv.org/abs/1404.5839}{{\tt[1404.5839]}}.

\bibitem{Anastasiou:2013srw}
C.~Anastasiou, C.~Duhr, F.~Dulat, and B.~Mistlberger,  {\em {Soft triple-real
  radiation for Higgs production at N3LO}}, {\em JHEP} {\bf 1307} (2013) 003,
\href{http://www.arXiv.org/abs/1302.4379}{{\tt[1302.4379]}}.

\bibitem{Shifman:1979eb}
A.~I. Vainshtein, M.~B. Voloshin, V.~I. Zakharov, and M.~A. Shifman,  {\em
  Low-energy theorems for {Higgs} boson couplings to photons}, {\em Yad. Fiz.}
  {\bf 30} (1979) 1368.
[{Sov}. J. {Nucl}. {Phys}. {\bf 30}, 711 (1979)].

\bibitem{Voloshin:1986tc}
M.~B. Voloshin,  {\em Once again about the role of gluonic mechanism in
  interaction of light {Higgs} boson with hadrons}, {\em Yad. Fiz.} {\bf 44}
  (1986) 738.
[{Sov}. J. {Nucl}. {Phys}. {\bf 44}, 478 (1986)].

\bibitem{Vainshtein:1980ea}
A.~I. Vainshtein, V.~I. Zakharov, and M.~A. Shifman,  {\em {Higgs} particles},
  {\em Usp. Fiz. Nauk} {\bf 131} (1980) 537.
[{Sov}. {Phys}. {Usp}. {\bf 23}, 429 (1980)].

\bibitem{Chetyrkin:1998un}
K.~Chetyrkin, B.~A. Kniehl, and M.~Steinhauser,  {\em {Decoupling relations to
  {${\cal O}\alpha_s^3$} and their connection to low-energy theorems}}, {\em
  Nucl.Phys.} {\bf B510} (1998) 61--87,
\href{http://www.arXiv.org/abs/hep-ph/9708255}{{\tt[hep-ph/9708255]}}.

\bibitem{Chetyrkin:1997iv}
K.~G. Chetyrkin, B.~A. Kniehl, and M.~Steinhauser,  {\em Hadronic {Higgs} decay
  to order {$\alpha_s^4$}}, {\em Phys. Rev. Lett.} {\bf 79} (1997) 353--356,
\href{http://arXiv.org/abs/hep-ph/9705240}{{\tt[hep-ph/9705240]}}.

\bibitem{Schroder:2005hy}
Y.~Schroder and M.~Steinhauser,  {\em {Four-loop decoupling relations for the
  strong coupling}}, {\em JHEP} {\bf 0601} (2006) 051,
\href{http://www.arXiv.org/abs/hep-ph/0512058}{{\tt[hep-ph/0512058]}}.

\bibitem{Chetyrkin:2005ia}
K.~Chetyrkin, J.~H. Kuhn, and C.~Sturm,  {\em {QCD decoupling at four loops}},
  {\em Nucl.Phys.} {\bf B744} (2006) 121--135,
\href{http://www.arXiv.org/abs/hep-ph/0512060}{{\tt[hep-ph/0512060]}}.

\bibitem{Ellis:1988xu}
R.~K. Ellis, I.~Hinchliffe, M.~Soldate, and J.~J. van~der Bij,  {\em {Higgs}
  decay to {$\tau^+\tau^-$}: A possible signature of intermediate mass {Higgs}
  bosons at the {SSC}}, {\em Nucl. Phys.} {\bf B297} (1988)
221.

\bibitem{Gehrmann:2011aa}
T.~Gehrmann, M.~Jaquier, E.~Glover, and A.~Koukoutsakis,  {\em {Two-Loop QCD
  Corrections to the Helicity Amplitudes for $H \to$ 3 partons}}, {\em JHEP}
  {\bf 1202} (2012) 056,
\href{http://www.arXiv.org/abs/1112.3554}{{\tt[1112.3554]}}.

\bibitem{Nogueira:1993ex}
P.~Nogueira,  {\em Automatic {Feynman} graph generation}, {\em J. Comput.
  Phys.} {\bf 105} (1993)
279--289.

\bibitem{Vermaseren:2000nd}
J.~Vermaseren,  {\em {New features of FORM}},
\href{http://www.arXiv.org/abs/math-ph/0010025}{{\tt[math-ph/0010025]}}.

\bibitem{vonManteuffel:2012np}
A.~von Manteuffel and C.~Studerus,  {\em {Reduze 2 - Distributed Feynman
  Integral Reduction}},
\href{http://www.arXiv.org/abs/1201.4330}{{\tt[1201.4330]}}.

\end{thebibliography}
\end{document}